# Ranking Triples using Entity Links in a Large Web Crawl

## The Chicory Triple Scorer at WSDM Cup 2017

Frank Dorssers
Radboud University
frank.dorssers@gmail.com

Arjen P. de Vries
Radboud University
arjen@cs.ru.nl

Wouter Alink
Spinque
wouter@spinque.com

Roberto Cornacchia
Spinque
roberto@spinque.com

## ABSTRACT

The Chicory Triple Scorer combines information derived from an entity's Wikipedia abstract with counts of entity mentions in a large annotated web crawl.

## 1. INTRODUCTION

This paper describes the participation of team Chicory in the Triple Ranking Challenge [2] of the WSDM Cup 2017 [7]. Our approach deploys a large collection of entity tagged web data to estimate the correctness of the relevance relation expressed by the triples, in combination with a baseline approach using Wikipedia abstracts following [1]. Relevance estimations are drawn from ClueWeb12 annotated by Google's entity linker, available publicly as the FACC1 dataset. Our implementation is automatically generated from a so-called 'search strategy' that specifies declaratively how the input data are combined into a final ranking of triples.

## 2. RELATED WORK

Bast et al. describe the challenges that have to be tackled for triple ranking and discuss how to create systems that deal with these problems [1]. They suggest a baseline method using Wikipedia that we have applied in our research as well.

Sawant and Chakrabarti discuss a similar problem, where they annotate natural language queries with structured knowledge [13]. Tonon et al. propose a method for entity ranking that is based on entity hierarchies extracted from structured knowledge bases like Freebase and DBpedia [14]. Given an entity and its textual context, they try to predict its possible types from the hierarchy depth and ancestry, where types with relevant ancestry receive a higher relevance score than types without.

## 3. APPROACH

We present our approach to address the triple ranking challenge; subsequently discussing the data used, the strategies for score prediction, and their implementation.

### 3.1 Data

Chicory's relevance scoring combines evidence from four data sources: the provided data (extended with Freebase identifiers), DBpedia, ClueWeb12 and FACC1. The information in these datasets can be linked together using Freebase [3] IDs, even if the Freebase API has ceased to exist. Ranking nationalities used an additional dictionary to relate countries to their nationalities.

#### 3.1.1 Provided data

The data released at the start of the challenge contains the following pieces of information:

- All possible people, nationalities and professions;
- knowledge bases for all possible combinations of people and nationalities and professions;
- relevance scores for part of the knowledge base;
- wiki sentences relevant to the provided entities.

Two attributes were missing for the approach described in this paper, and the following adjustments were made. First, Freebase IDs were added to nationalities and professions where possible. Second, the provided nationalities (which currently listed countries), have been extended with their actual nationalities (for example, adding 'Dutch' to 'Netherlands' and 'American' to 'United States of America'), using an online dictionary.[1]

#### 3.1.2 DBpedia

DBpedia [8] is an effort by Leipzig University, University of Mannheim and OpenLink software to provide a version of Wikipedia in a linked data format based on RDF. RDF data can be queried using SPARQL [11], giving efficient access to specific properties, relations, and information about given entities.

DBpedia is currently used to look up abstracts for persons and to find Freebase IDs for entities. For performance reasons, the RDF graphs for the persons considered in the challenge were scraped instead of running SPARQL queries to extract just the abstract.

The process of adding Freebase IDs to professions and nationalities using DBpedia consists of two steps:

1. Automated lookup on DBpedia using SPARQL;
2. manual verification of IDs found.

For the first step, a SPARQL query checks the resource page of a given entity for the `owl:sameAs` relationship. This relationship links to multiple different external data sources, however, in this case only the Freebase relationship is used.

Specific entities may not have a corresponding DBpedia page; consider for example the profession 'activist'. While 'activist' is not available, 'activism' is; and, the 'activism' page links back to 'activist' through the `dbo:wikiPageRedirects` relation.

---

[1] See http://www.esldesk.com/vocabulary/countries, last accessed January 24th, 2017.

Automatically retrieved Freebase IDs are manually verified against available Wikidata Freebase IDs and checked against FACC1 to make sure they are actually present. Unused Freebase IDs are removed and Freebase IDs are manually looked up for professions or nationalities that do not have one at this point.

### 3.1.3 ClueWeb12

The ClueWeb12 dataset is a partial crawl of the World Wide Web, consisting of 733,019,372 English webpages which were crawled between February 10, 2012 and May 10, 2012. More details about the data are available on their webpage.[2]

Currently only the ClueWeb12 document IDs are used in our method. Future extensions of the approach will however deploy the actual text information available on these pages.

### 3.1.4 FACC1

We also use the Freebase Annotations of the Clueweb Corpora, v1 dataset [6], referred to as FACC1. These annotations were produced automatically by Google, with an entity linker tuned for high precision.

Consider an example annotation for entity 'George Clooney':
`clueweb12-0506wb-91-00045, UTF-8, George Clooney, 10651, 10665, 0.995516, 0.000126, /m/014zcr`

This contains several interesting pieces of information: the document ID in ClueWeb12, the encoding, the exact string that was found in the document, byte-offsets of start and end of the entity mention, the probability that this piece of text (including context) corresponds to this Freebase entity, the probability that the context (excluding the exact text) relates to the Freebase entity, and then, finally, the actual Freebase ID.

These annotations open up all kind of possibilities. You can extract context about entity mentions, find what entities occur together in a document, give these mentions a weight based on the provided probability, and so forth. The primary use in our current approach is to identify co-occurrences of people and nationalities or professions.

The FACC1 set consists of annotations of 456,498,584 documents, with approximately 13 mentions per document, resulting in almost 6 million annotations. To speed up the pipeline, the annotations went through two pre-processing steps before actually being used. In the first step, annotations which were not about a profession, nationality or person were filtered out. In the second step, documents without co-occurrences were removed, i.e., every remaining document contains an annotation of a person as well as either a profession or a nationality. The resulting set contains 771,813,525 annotations.

## 3.2 Strategies

The Chicory Triple Ranker uses two strategies to predict relevance scores, which are combined to provide the final score.

### 3.2.1 Abstract location

The *abstract location strategy* is a baseline method, deriving its score from the location of the profession or nationality mention in the abstract, following one of the baselines proposed in [1]. A nationality or profession is assigned a relevance score of seven if it is the first mention of that type in the text of the abstract on the person's Wikipedia page, zero otherwise.

### 3.2.2 FACC1 count

The *FACC1 count strategy* bases its relevance score on co-occurrences of people and nationalities or professions within a single document.

An important issue to note is that while large variances between person mentions do not matter (for example John appears 50.000 times, Pete only 50), they do matter for professions and nationalities. The profession 'politician' occurs much more frequently than 'harpsichordist'. Thus, co-occurrences of 'politician' and a specific person are much more likely. This is currently addressed by taking the log of all occurrences, but different weighting strategies should be considered in future work, for example by normalization with the total number of occurrences.

The results after taking the log are normalized to a 0-7 scale.

### 3.2.3 Combining strategy result

Many different ways of combining results from the different strategies have been proposed in the scientific literature on score and rank fusion. We used the straightforward combination of taking the maximum of the relevance scores under consideration, or a default score if both values were zero (set to 4 in our experiments for the challenge).

## 3.3 Implementation

This section describes how the previously described strategies are implemented. The goal is to create an implementation which was independent of the triple type and would be able to predict types even outside of the provided set.

### 3.3.1 Modeling strategies

Strategies have been modeled using Spinque Desk [5], a graphical editor which allows for designing search strategies declaratively, by connecting building blocks (see Figure 1). Each building block encapsulates code snippets for the implementation of basic tasks including selection, graph navigation, and ranking. Upon execution, these code snippets are combined and the resulting program optimized and finally converted to SQL queries. Score computation is also translated into SQL, with the help of Probabilistic Relational Algebra [12] as an intermediate abstraction layer.

### 3.3.2 Database engine

The previously described SQL queries are executed on the MonetDB column store [4]. MonetDB is optimized for analytical processing and provides robustness for the data. Together with it being a column store and general optimizations makes MonetDB uniquely suited for information retrieval tasks. See the paper by Mühleisen et al. for more information on why column stores are interesting for rapid prototyping [9] and the paper by Cornacchia et al. why a database is used in this context [5].

### 3.3.3 Python

A custom Python framework is used to interface with MonetDB and run the generated SQL queries based on PRA. This framework has been set up with usability in mind, allowing for easy switching of different strategies and approaches. It handles all the in- and output required for this challenge: the initial command line input and the correctly formatted output for the provided `evaluator.py`.

### 3.3.4 Runtime

Running the complete implementation on TIRA [10] requires just a MonetDB database server on the virtual machine, and a driver program to issue the generated SQL queries (which has been done using Python and the MonetDB library).

---

[2]See http://lemurproject.org/clueweb12/index.php, last accessed January 24th, 2017.

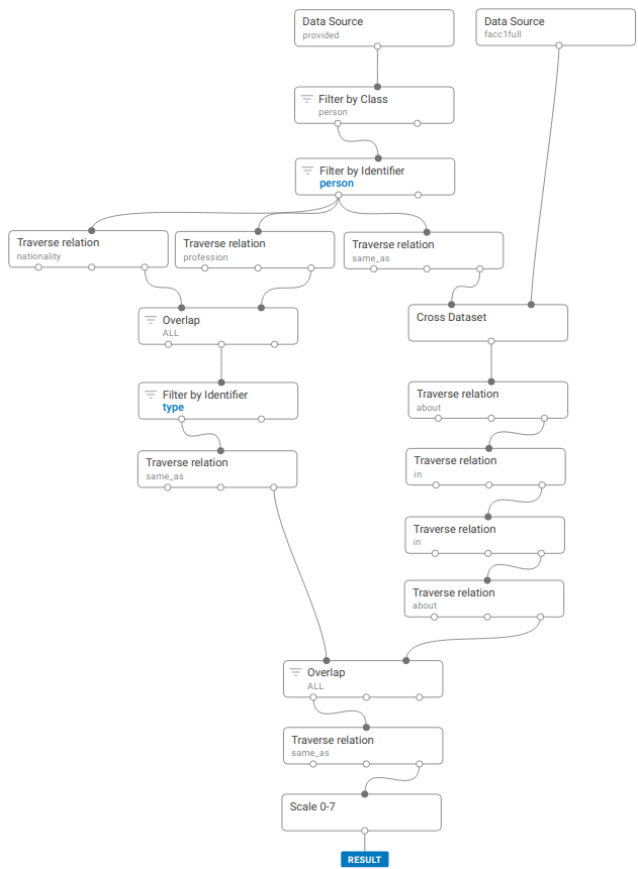

**Figure 1: Strategy for co-occurrence counting in FACC1**

### 3.3.5 FACC1 count implementation

Figure 1 shows the implementation of the FACC1 count strategy in Spinque Desk and this section will take a closer look at what these blocks do and how they form the eventual strategy.

The datasets used for these tasks are stored in their own tables with RDF formatting. For the FACC1 data this means that the single line as seen before in section 3.1.4 has been turned into three classes (the document, the annotation and the entity it is about) and two relationships (what document the annotation is in and what entity it is about). The files containing the persons, professions and nationalities have been combined with the knowledge bases. This new format results in four classes (Freebase, person, nationality and profession), and two types of relationships (connecting the entities to Freebase objects and connecting persons to their respective nationalities and professions.

The strategy shown in Figure 1 is read from top to bottom, takes two inputs and produces one output. The inputs are the type that is currently being predicted, profession or nationality, and the person for whom it is being predicted. These two inputs can be seen as blue and bolded text in the `Filter by Identifier` blocks. This strategy produces a single output, namely 0-7 scores for all professions or nationalities that are relevant according to the knowledge base and that are present in the FACC1 data. Any missing entities are handled by Python, which assumes that the score is zero.

The strategy can be divided into a left and right side, which are combined to generate the actual results. The left side produces all Freebase IDs for all professions or nationalities that are relevant for the provided person, while the right side selects all relevant data from FACC1 based on the input and generates counts for each entity.

The left side starts by loading all 'provided' data, filtering out anything that is not a person and then selecting the one that has been provided as input. The two traversals on the left produce all relevant nationalities and professions, which are then combined, followed by a filter which leaves only objects of the type (profession or nationality) that has been provided as input. The final traversal on this side gives all the connected Freebase IDs for the remaining entities.

The right side starts with a dataset crossover, where the Freebase ID of the selected person from the left side is used to select the correct Freebase object in the FACC1 dataset. The following traversals go back and forth over Freebase, document and annotation, starting with the Freebase object for the current person. Spinque desk tries to minimize cluttering during the design phase, so several options, like what way the traversal goes, or whether or not the results are aggregated, are not immediately visible on the block itself, but become visible after selecting it. The first traversal, over 'about', produces all annotations that are about this person. The second traversal, over 'in', gives all documents in which these annotations are present. Traversing back the other way over 'in' gives all annotations that are in these documents, so not just the annotations about the person. The last of the four traversals, over 'about', is again back the other way, now producing all Freebase IDs that these annotations are about, thus also all Freebase IDs that occur together with the person. An important thing to note is that this last traversal also does an aggregation, so it produces counts for each Freebase ID how often it occurs.

The final steps in this strategy are checking the intersection between the relevant Freebase IDs from the left side and all the Freebase IDs from the right side, leaving only counts for Freebase IDs that are relevant to the user. The final traversal then turns these Freebase IDs back into the actual entities, after which the counts are scaled between 0 and 7 and they are returned.

## 4. EVALUATION RESULTS

Bast, Buchhold and Haussmann provide a complete overview of scores by all approaches in this challenge [2]. Here, we discuss in detail our experimental findings and their implications.

### 4.1 Challenge results

The approach described in this paper achieved an accuracy of 0.63, an average score difference of 1.97 and a Kendall tau score of 0.35. These measures report the weighted average of the results for the professions and nationalities. Rerunning the algorithm on the released test data shows a (slightly) better performance on nationalities than professions, with an accuracy of 0.66, an average score difference of 1.82 and an average Kendall's Tau of 0.38 (as opposed to 0.62, 2.03 and 0.34).

### 4.2 Additional experiments

The results submitted for the challenge were achieved with a default relevance score of 4, used whenever the strategies did not find any evidence. Due to time constraints and technical limitations, the submitted runs used only a sample of the filtered FACC1 data (51.2 million annotations to be precise), as opposed to the 771 million filtered annotations that could be used.

The choice of default value had a significant effect on the performance of the strategies in this setting; using a default value of 0 results in scores of only 0.57, 2.34 and 0.38 for nationalities and 0.50, 2.81 and 0.34 for professions. The large decrease of scores with a sub-optimal default value can be explained by a lack of evi-dence for many triples and sparseness of the data. Out of the 197 nationalities in the test data, the strategies were not able to find evidence for 44 triples, for the professions this was 239 out of 513.

Post challenge, an additional local run has been performed on the complete data, resulting in scores of 0.63, 1.94 and 0.38 for nationalities and 0.62, 2.13 and 0.35 for professions, using 4 as the default value. Changing the default value from 4 to 0 now has almost no effect on the scores for nationality relevance scores, while it still has quite an influence on the profession scores. This indicates that the sparseness of the nationality data may have been resolved, while for professions missing evidence remains an issue. Indeed, this can be verified by checking the data, showing that only five nationalities receive the default score, while professions still suffer from missing data for 153 cases.

Table 1: Evaluation results (professions).

| Method | Acc | Asd | Kendall-$\tau$ |
| --- | --- | --- | --- |
| Abstracts | 0.47 | 3.04 | 0.35 |
| Partial counts | 0.41 | 3.31 | 0.42 |
| Full counts | 0.44 | 3.20 | 0.44 |
| Both (partial, 4) | 0.62 | 2.03 | 0.34 |
| Both (partial, 0) | 0.50 | 2.81 | 0.34 |
| Both (full, 4) | 0.62 | 2.13 | 0.35 |
| Both (full, 0 | 0.51 | 2.73 | 0.34 |

Table 2: Evaluation results (nationalities).

| Method | Acc | Asd | Kendall-$\tau$ |
| --- | --- | --- | --- |
| Abstracts | 0.48 | 2.87 | 0.34 |
| Partial counts | 0.50 | 2.77 | 0.51 |
| Full counts | 0.63 | 2.12 | 0.51 |
| Both (partial, 4) | 0.66 | 1.82 | 0.38 |
| Both (partial, 0) | 0.57 | 2.34 | 0.38 |
| Both (full, 4) | 0.63 | 1.94 | 0.38 |
| Both (full, 0) | 0.63 | 2.01 | 0.38 |

Tables 1 and 2 show the performance metrics for the submitted runs compared to alternative, post challenge runs. Run labels indicate the setting, where (partial, 4) indicates a run using only a part of FACC1 with a default value of 4. We include three single strategy runs, using the abstracts and the partial or full FACC1 data.

### 4.3  Results analysis

The accuracy scores for the professions show that all single strategies score below 0.5; combining the approaches improves the results in all cases, indicating that the two strategies provide different evidence. A combination strategy is also best for nationalities, except for the strategies using the full FACC1 data; this manages to return information for almost any combination provided, so the abstract strategy contributes less to the result.

We have identified a number of aspects for improvement in the current usage of data as well as the overall approach.

While Freebase IDs exist for many professions, not all of them do actually have one. This results in either missing data or less accurate data; take for example 'book editor' and 'film editor', where only a more generic 'editor' Freebase ID exists. In this case, both professions were given the same ID, as this was preferred over completely losing this information.

Table 3: Comparison of predicted vs test scores for the profession test set using full count strategy

| | Predicted scores | | | | | | | |
| --- | --- | --- | --- | --- | --- | --- | --- | --- |
| True scores | 0 | 1 | 2 | 3 | 4 | 5 | 6 | 7 |
| 0 | 17 | 1 | 2 | 11 | 15 | 7 | 5 | 10 |
| 1 | 14 | 1 | 2 | 4 | 6 | 7 | 9 | 6 |
| 2 | 18 | 2 | 1 | 7 | 9 | 7 | 2 | 4 |
| 3 | 16 | 2 | 0 | 1 | 6 | 6 | 5 | 7 |
| 4 | 26 | 1 | 1 | 5 | 5 | 4 | 4 | 10 |
| 5 | 25 | 1 | 2 | 5 | 3 | 4 | 3 | 14 |
| 6 | 29 | 4 | 2 | 5 | 5 | 7 | 5 | 14 |
| 7 | 55 | 1 | 3 | 3 | 5 | 8 | 3 | 41 |

Table 4: Comparison of predicted vs test scores for the nationality test set using full count strategy

| | Predicted scores | | | | | | | |
| --- | --- | --- | --- | --- | --- | --- | --- | --- |
| True scores | 0 | 1 | 2 | 3 | 4 | 5 | 6 | 7 |
| 0 | 1 | 0 | 0 | 1 | 0 | 2 | 3 | 1 |
| 1 | 0 | 0 | 0 | 0 | 1 | 1 | 5 | 6 |
| 2 | 0 | 0 | 0 | 0 | 0 | 2 | 2 | 6 |
| 3 | 0 | 0 | 2 | 1 | 0 | 1 | 4 | 11 |
| 4 | 1 | 0 | 0 | 0 | 1 | 2 | 5 | 18 |
| 5 | 1 | 0 | 0 | 0 | 2 | 3 | 3 | 19 |
| 6 | 0 | 0 | 0 | 1 | 2 | 4 | 8 | 23 |
| 7 | 5 | 0 | 0 | 0 | 2 | 1 | 12 | 34 |

A different challenge with regard to Freebase ID lookups is the question how to find IDs of entities that are on the same 'semantic' level of detail. Consider the two professions 'activist' and 'politician', both occurring in the challenge data. DBpedia redirects the 'activist' page to 'activism', while 'politician' does has its own page. As a result, the Freebase ID for 'activism' is used instead of the one for 'activist', however, 'activism' has a different type of meaning than 'politician' (i.e., a person's activity versus a person's role). In this case, you can either use it *as is* and accept the semantic difference, or lose information by going up a level on both sides (i.e., taking the Freebase ID for 'politics' instead of 'politician'). The latter would however create a different problem, because selecting the higher-level entity might not be obvious for every profession. For this reason, it was decided to use the Freebase ID for the entity as close as possible to the term of the provided profession.

Tables 3 and 4 show confusion matrices for the scores generated when solely using the full counts for predictions. These matrices indicate that the lack of co-occurrences in the FACC1 data is especially harmful for the profession predictions, where 183 out of the 200 are wrongly scored as zero. Predictions for nationalities do not suffer that much from lack of data, as the data is quite complete. This was already mentioned as being the cause as to why adding abstract information does not increase accuracy. A large issue however with the nationality data is that this table shows that too many triples are ranked seven, only 35 out of the 126 ranked as seven actually should be seven.

Another question is how to improve the handling of missing co-occurrences of person and nationality or profession Freebase IDs in FACC1. A lack of information has to be compensated for, which is currently done using the baseline abstract strategy. A better back-off solution might use additional data from ClueWeb12, for example by analyzing the text in the context of person mentions.

## 5. CONCLUSION

We presented a straightforward, data-driven approach to the Triple Ranking Challenge. Effectiveness of the method is low in comparison to the results of other teams. Results did improve by processing the complete FACC1 corpus, but not with a large enough margin to be competitive. We conclude that especially the weighting of evidence needs to be improved. Additional analysis is necessary before we can firmly establish if the cause of the low performance can be attributed to erroneous and missing data, or that a different way of processing the data would be necessary for a better performance on the task.

With respect to the implementation, we found that we are currently limited by the scalability of the platform deployed for processing the underlying web data. A version of the runtime engine that scales to the size of ClueWeb12 would be desirable. We may need to lift parts of the query processing to a cluster (e.g. running the Spark platform).